\documentstyle[11pt,psfig,twoside]{article}

\begin{document}

\newcommand{\dd}{\,{\rm d}}
\newcommand{\ie}{{\it i.e.},\,}
\newcommand{\etal}{{\it et al.\ }}
\newcommand{\eg}{{\it e.g.},\,}
\newcommand{\cf}{{\it cf.\ }}
\newcommand{\vs}{{\it vs.\ }}
\newcommand{\zdot}{\makebox[0pt][l]{.}}
\newcommand{\up}[1]{\ifmmode^{\rm #1}\else$^{\rm #1}$\fi}
\newcommand{\dn}[1]{\ifmmode_{\rm #1}\else$_{\rm #1}$\fi}
\newcommand{\upd}{\up{d}}
\newcommand{\uph}{\up{h}}
\newcommand{\upm}{\up{m}}
\newcommand{\ups}{\up{s}}
\newcommand{\arcd}{\ifmmode^{\circ}\else$^{\circ}$\fi}
\newcommand{\arcm}{\ifmmode{'}\else$'$\fi}
\newcommand{\arcs}{\ifmmode{''}\else$''$\fi}
\newcommand{\MS}{{\rm M}\ifmmode_{\odot}\else$_{\odot}$\fi}
\newcommand{\RS}{{\rm R}\ifmmode_{\odot}\else$_{\odot}$\fi}
\newcommand{\LS}{{\rm L}\ifmmode_{\odot}\else$_{\odot}$\fi}

\newcommand{\Abstract}[2]{{\footnotesize\begin{center}ABSTRACT\end{center}
\vspace{1mm}\par#1\par
\noindent
{~}{\it #2}}}

\newcommand{\TabCap}[2]{\begin{center}\parbox[t]{#1}{\begin{center}
  \small {\spaceskip 2pt plus 1pt minus 1pt T a b l e}
  \refstepcounter{table}\thetable \\[2mm]
  \footnotesize #2 \end{center}}\end{center}}

\newcommand{\TableSep}[2]{\begin{table}[p]\vspace{#1}
\TabCap{#2}\end{table}}

\newcommand{\FigCap}[1]{\footnotesize\par\noindent Fig.\  %
  \refstepcounter{figure}\thefigure. #1\par}

\newcommand{\TableFont}{\footnotesize}
\newcommand{\TableFontIt}{\ttit}
\newcommand{\SetTableFont}[1]{\renewcommand{\TableFont}{#1}}

\newcommand{\MakeTable}[4]{\begin{table}[htb]\TabCap{#2}{#3}
  \begin{center} \TableFont \begin{tabular}{#1} #4 
  \end{tabular}\end{center}\end{table}}

\newcommand{\MakeTableSep}[4]{\begin{table}[p]\TabCap{#2}{#3}
  \begin{center} \TableFont \begin{tabular}{#1} #4 
  \end{tabular}\end{center}\end{table}}

\newenvironment{references}%
{
\footnotesize \frenchspacing
\renewcommand{\thesection}{}
\renewcommand{\in}{{\rm in }}
\renewcommand{\AA}{Astron.\ Astrophys.}
\newcommand{\AAS}{Astron.~Astrophys.~Suppl.~Ser.}
\newcommand{\ApJ}{Astrophys.\ J.}
\newcommand{\ApJS}{Astrophys.\ J.~Suppl.~Ser.}
\newcommand{\ApJL}{Astrophys.\ J.~Letters}
\newcommand{\AJ}{Astron.\ J.}
\newcommand{\IBVS}{IBVS}
\newcommand{\PASP}{P.A.S.P.}
\newcommand{\Acta}{Acta Astron.}
\newcommand{\MNRAS}{MNRAS}
\renewcommand{\and}{{\rm and }}
\section{{\rm REFERENCES}}
\sloppy \hyphenpenalty10000
\begin{list}{}{\leftmargin1cm\listparindent-1cm
\itemindent\listparindent\parsep0pt\itemsep0pt}}%
{\end{list}\vspace{2mm}}

\def\TYLDA{~}
\newlength{\DW}
\settowidth{\DW}{0}
\newcommand{\dw}{\hspace{\DW}}

\newcommand{\refitem}[5]{\item[]{#1} #2%
\def\REFARG{#3}\ifx\REFARG\TYLDA\else, {\it#3}\fi
\def\REFARG{#4}\ifx\REFARG\TYLDA\else, {\bf#4}\fi
\def\REFARG{#5}\ifx\REFARG\TYLDA\else, {#5}\fi.}

\newcommand{\Section}[1]{\section{#1}}
\newcommand{\Subsection}[1]{\subsection{#1}}
\newcommand{\Acknow}[1]{\par\vspace{5mm}{\bf Acknowledgements.} #1}
\pagestyle{myheadings}

\def\thefootnote{\fnsymbol{footnote}}
\begin{center}
{\Large\bf The Optical Gravitational Lensing Experiment.\\
\vskip3pt
Cepheids in the Magellanic Clouds.\\
\vskip3pt
II. Single-Mode Second Overtone Cepheids\\
\vskip3pt
in the Small Magellanic Cloud\footnote{Based on  observations obtained
with the 1.3~m Warsaw telescope at the Las Campanas  Observatory of the
Carnegie Institution of Washington.}}
\vskip1cm
{\bf
A.~~U~d~a~l~s~k~i$^1$,~~I.~~S~o~s~z~y~{\'n}~s~k~i$^1$,
~~M.~~S~z~y~m~a~{\'n}~s~k~i$^1$,~~M.~~K~u~b~i~a~k$^1$,
~~G.~~P~i~e~t~r~z~y~\'n~s~k~i$^1$,
~~P.~~W~o~\'z~n~i~a~k$^2$,~~ and~~K.~~\.Z~e~b~r~u~\'n$^1$}
\vskip3mm
{$^1$Warsaw University Observatory, Al.~Ujazdowskie~4, 00-478~Warszawa, Poland\\
e-mail: (udalski,soszynsk,msz,mk,pietrzyn,zebrun)@sirius.astrouw.edu.pl\\
$^2$ Princeton University Observatory, Princeton, NJ 08544-1001, USA\\
e-mail: wozniak@astro.princeton.edu}
\end{center}

\Abstract{We present a sample of 13 candidates for single-mode second
overtone  Cepheids discovered in the 2.4 square degree area of the SMC
bar during the OGLE  microlensing search. All candidates passed four
photometric tests: analysis of  Fourier parameters of the light curve,
$R_{21}$ and $\phi_{21}$, position in  the period-luminosity diagram and
location in the color-magnitude diagram. 

The stars form a very homogeneous group of objects with very similar
light  curves of low amplitude and almost sinusoidal shape. Their color
indices ${B-V}$ and  ${V-I}$ are also very similar and the objects
determine the observational blue edge of the instability strip of the
first overtone Cepheids. This is the first firm  sample of the
single-mode second overtone Cepheids detected so far.}{~}

\Section{Introduction}
The existence of second overtone pulsations in the Cepheid variable
stars was  predicted by Stobie (1969). For years this was only a
theoretical possibility  as no such object has been found. The first
indication of existence of second  mode pulsations was the discovery of
double-mode Cepheid CO~Aur (Mantegazza  1983) which is believed to
pulsate simultaneously in the first and second  overtones. This is the
only Galactic object of that kind. Another Galactic  object, HR~7308,
was proposed to be a single-mode second overtone Cepheid  candidate
(Burki \etal 1986). 

The major breakthrough in this field occurred when huge photometric
databases  collected during the microlensing search programs were
analyzed. Two targets  of microlensing searches, the Large and Small
Magellanic Clouds are ideal  objects for Cepheid study possessing huge
population of these stars,  additionally  located at approximately the
same distance. 

The MACHO team reported discovery of 45 double-mode Cepheids pulsating
in the  first and second overtones in the LMC (Alcock \etal 1999) and 20
such objects  in the SMC (Alcock \etal 1997). The EROS team discovered
additional sample of  7 such objects in the SMC (Beaulieu \etal 1997). 

The MACHO team also undertook a search for Cepheids pulsating solely in
the  second overtone (Alcock \etal 1999). Based on Fourier decomposition
of light  curves of double-mode Cepheids pulsating in the first and
second overtones  they characterized properties of second overtone
pulsations and provided  constraints on Fourier parameters allowing
separation of the second overtone  pulsators from Cepheids of other
type. They found one potential candidate for  single-mode second
overtone Cepheid in their sample of about 1400 Cepheids in  the LMC.
Also theoretical modeling of second overtone pulsations in Cepheids  by
Antonello and Kanbur (1997) confirmed qualitatively the MACHO team
results. 

The large amount of {\it BVI} photometric data of the SMC collected
during the  Optical Gravitational Lensing Experiment (OGLE) provides an
unique material  for analysis of variable stars in this galaxy. Lower
metallicity of the SMC  as compared to the LMC makes this object
particularly attractive for searching  for second overtone pulsating
Cepheids because such stars should be more  numerous in low metallicity
environment (Antonello and Kanbur 1997). Indeed,  the largest sample of
double-mode Cepheids (23 pulsating in the fundamental  mode and first
overtone and 70 pulsating in the first and second overtones)  discovered
so far in one galaxy was reported by the OGLE team (Udalski \etal 
1999). 

In this paper we describe the search for pure  second overtone
pulsations, \ie  single-mode second overtone Cepheids. The search among
about 2300 Ce\-pheids  detected in the 2.4 square degree area in the
central bar of the SMC led to  discovery of 13 candidates -- the first
firm sample of single-mode second  overtone pulsating Cepheids. 

\Section{Observations}
All observations presented in this paper were carried out during the
second  phase of the OGLE experiment with the 1.3-m Warsaw telescope at
the Las  Campanas Observatory, Chile, which is operated by  the Carnegie
Institution of  Washington. The telescope was equipped with the "first
generation" camera with   a SITe ${2048\times2048}$ CCD detector working
in the drift-scan mode. The  pixel size was 24~$\mu$m giving the 0.417
arcsec/pixel scale. Observations of  the SMC were performed in the
"slow" reading mode of the CCD detector with the  gain 3.8~e$^-$/ADU and
readout noise about 5.4~e$^-$. Details of the  instrumentation setup can
be found in Udalski, Kubiak and Szyma{\'n}ski  (1997). 

Observations of the SMC started on June~26, 1997. As the microlensing
search   is planned to last for a few years, observations of selected
fields will be  continued during the following seasons. In this paper we
present data  collected up to March~4, 1998. Observations were obtained
in the standard {\it  BVI}-bands  with majority of measurements made in
the {\it I}-band. 

Photometric data collected during the first observing season of the SMC
for  11 fields (SMC$\_$SC1--SMC$\_$SC11) covered about 2.4 square degree
of the  central parts of the SMC and were used to construct the {\it
BVI} photometric  maps of the SMC (Udalski \etal 1998). The reader is
referred to that paper for  more details about methods of data
reduction, tests on quality of photometric  data, astrometry, location
of observed fields etc. 

\Section{Search for Second Overtone Cepheids}
In a search for single-mode second overtone Cepheid candidates in our
sample of  SMC Cepheids we followed strategy outlined by the MACHO group
(Alcock \etal  1999), who used the double-mode Cepheids pulsating in the
first and second  overtones to characterize the most important
properties of second overtone  pulsations. Based on analysis of
parameters $R_{21}$ and $\phi_{21}$ of  Fourier decomposition of light
curves of double-mode Cepheids they concluded  that the second overtone
pulsations should be small amplitude, almost  sinusoidal light
variations. They constrained area in the $R_{21}$ \vs $\log  P$ and
$\phi_{21}$ \vs $\log P$ diagrams where the second overtone pulsations 
could be distinguishable from the first overtone and fundamental mode
Cepheids  and found one possible candidate for second overtone Cepheid
in their sample  of about 1400 LMC Cepheids. Similar conclusions on
properties of second  overtone single-mode pulsations were drawn by
Antonello and Kanbur (1997) from  theoretical models. 

With larger sample of double-mode Cepheids found by the OGLE team in the
SMC  (Udalski \etal 1999), we were in position to perform independent
analysis of  properties of second overtone pulsations in double-mode
Cepheids. Appendix~C  in Udalski \etal (1999) presents light curves of
double-mode Cepheids  decomposed to both pulsation modes (column~3 --
second overtone pulsations,  column~4 -- first overtone pulsations). It
is striking that the light curves  of the second overtone pulsations
look almost identical -- they are indeed small  amplitude, almost
sinusoidal light variations. More quantitative conclusions  can be drawn
from Fig.~4 of Udalski \etal (1999) which presents $R_{21}$ \vs  $\log
P$ and $\phi_{21}$ \vs $\log P$ diagrams for Cepheids in the SMC with 
positions of double-mode first and second overtone Cepheids. It is clear
that  in most cases the $R_{21}$ values of the second overtone
pulsations are  small (${R_{21}<0.2}$; if this parameter is not
statistically significant  $R_{21}$ is shown as zero, meaning sinusoidal
light curve). The area limited  by ${R_{21}<0.2}$ and ${\log P<0}$ is
the best region where the second  overtone pulsations can be
distinguished from the first overtone Cepheids  which practically do not
populate that area. Similarly, the $\phi_{21}$ values  of the second
overtone Cepheids mostly populate regions above the sequence of  the
first overtone Cepheids in the $\phi_{21}$ \vs $\log P$ diagram. This is
another way of distinguishing objects pulsating in these modes. Thus,
our  results are in good agreement with Alcock \etal (1999) study of LMC
double-mode Cepheids. $R_{21}$ and $\phi_{21}$ parameters of the Fourier
decomposition of the Cepheid light curve provide two important
constraints  which single-mode second overtone Cepheid candidates should
fulfill. 

The shape of the light curve is not sufficient to make sound
identification of  the potential single-mode second overtone Cepheids.
It is well known that many  variable objects reveal small amplitude
almost sinusoidal light variations.  The best example are ellipsoidal
binary variable stars in which changing  aspects during orbital motion
of distorted components result in low amplitude  sinusoidal variations.
Another examples are binary stars with reflex effect,  spotted stars
etc. Fortunately, the Cepheid variable stars occupy a well  defined
region in the color-magnitude diagram (CMD), practically not populated 
by stars of other type. Thus, by constraining colors and luminosities of
potential candidates in the CMD diagram  one can get rid of most of 
contaminating stars: ellipsoidal variable stars which are usually hot
main  sequence objects, spotted stars -- usually late type giants or
main sequence  stars of low luminosity etc. 

Another constraint can be imposed by analyzing the period of variability
and  the mean luminosity of candidates. It is well known that the period
of the  first overtone pulsations is in double-mode Cepheids by factor
of 0.74  shorter than that of the fundamental mode. Thus, the single
mode first  overtone pulsators should be shifted by ${|\log0.74|=0.131}$
to the left in  the period-luminosity (PL) diagram ($I$ \vs $\log P$).
Indeed, clear  separation of the first overtone and fundamental mode
Cepheids is observed in  PL diagrams of Cepheids from the LMC (Alcock
\etal 1999) and SMC (see below). 

By analogy the period of the second overtone Cepheids is by factor of
0.805  shorter than the period of the first overtone variations in
double-mode  Cepheids. Therefore one can expect that the single-mode
second overtone  Cepheids should be shifted by ${|\log0.805|=0.094}$ to
the left from the first  overtone Cepheid sequence in the PL diagram.
The shift is smaller than in the  case of the fundamental mode and first
overtone Cepheids. Natural width of the  sequence of first overtone
Cepheids, additionally broadened by differential  reddening and
observational errors, makes this constraint not very strong, if 
considered alone. Nevertheless, it is an important part of the 
multi-constraint test described above. 

\Section{Results of Search}
We started our search from the analysis of Fourier parameters of light
curve  decomposition of all 2300 Cepheids detected in the SMC. We
selected candidates  with $R_{21}$ in the regions of $R_{21}$ \vs $\log
P$ diagram where the second  overtone pulsations in double-mode Cepheids
are located. Then we  eliminated all objects for which $\phi_{21}$
parameter fell in the first  overtone pulsation sequence in the
$\phi_{21}$ \vs $\log P$ diagram. 

In the next step we limited ourselves to the analysis of Cepheids for
which  {\it VI}-band observations were available. Only stars from the
following part  of the CMD diagram were considered: ${I<18.5}$~mag,
${0.25<(V-I)<1.3}$~mag.  Finally, we checked positions of our all second
overtone Cepheid candidates in  the PL diagram. Some of them were
located in the first overtone PL sequence  and as such they were
rejected from the final list. 

As a test we performed similar Fourier analysis of the remaining
variable star  candidates discovered in the OGLE variable stars search.
Some candidates  passed $R_{21}$ and $\phi_{21}$ criteria but, as we
expected, they turned out  to be ellipsoidal variable stars located in
the upper part of the main  sequence with ${V-I<0.2}$. 

%\vspace*{-12pt}
\tabcolsep=2pt
\renewcommand{\TableFont}{\scriptsize}
\MakeTable{c@{\hspace{-1pt}}rccccccccc}{12.5cm}{Single-mode second overtone 
Cepheid candidates in the SMC}
{\hline
\noalign{\vskip2pt}
Field       & Star No. & RA(J2000) & DEC(J2000) & $P$ & $R_{21}$ & $\phi_{21}$ & $I$ & ${B-V}$ & ${V-I}$ & Remarks\\
            &          &           &            & [days] &       &             &[mag]& [mag]   & [mag]   & \\
\noalign{\vskip2pt}
\hline
\noalign{\vskip2pt}
SMC$\_$SC3  & 193393 & 0\uph44\upm56\zdot\ups96 & $-73\arcd19\arcm37\zdot\arcs9$ & 0.63056 & 0.000 &  --   & 15.946 & 0.115 & 0.362 &uncertain\\
SMC$\_$SC4  &  22650 & 0\uph46\upm00\zdot\ups23 & $-73\arcd10\arcm32\zdot\arcs3$ & 0.78255 & 0.195 & 4.581 & 16.535 & 0.328 & 0.452 &\\
SMC$\_$SC4  &  71618 & 0\uph46\upm33\zdot\ups19 & $-73\arcd07\arcm45\zdot\arcs4$ & 0.70772 & 0.116 & 4.753 & 16.682 & 0.279 & 0.432 &\\
SMC$\_$SC4  &  97000 & 0\uph46\upm32\zdot\ups54 & $-72\arcd40\arcm31\zdot\arcs0$ & 0.70837 & 0.145 & 4.787 & 16.762 & 0.311 & 0.467 &\\
SMC$\_$SC4  & 167294 & 0\uph47\upm51\zdot\ups99 & $-73\arcd13\arcm33\zdot\arcs2$ & 1.12408 & 0.000 &  --   & 15.982 & 0.221 & 0.536 &\\
SMC$\_$SC4  & 179239 & 0\uph48\upm29\zdot\ups24 & $-73\arcd01\arcm16\zdot\arcs1$ & 0.63053 & 0.000 &  --   & 16.838 & 0.242 & 0.423 &\\
SMC$\_$SC5  & 235321 & 0\uph50\upm10\zdot\ups09 & $-72\arcd41\arcm51\zdot\arcs1$ & 0.73759 & 0.124 & 2.063 & 16.743 & 0.284 & 0.433 &\\
SMC$\_$SC6  &     93 & 0\uph52\upm04\zdot\ups47 & $-73\arcd24\arcm43\zdot\arcs2$ & 0.92159 & 0.000 &  --   & 16.176 & 0.250 & 0.344 &\\
SMC$\_$SC6  &  17477 & 0\uph51\upm57\zdot\ups62 & $-73\arcd14\arcm47\zdot\arcs9$ & 0.57225 & 0.133 & 2.414 & 16.756 & 0.113 & 0.416 &\\
SMC$\_$SC6  &  35769 & 0\uph51\upm26\zdot\ups26 & $-73\arcd02\arcm58\zdot\arcs0$ & 0.59371 & 0.274 & 2.976 & 16.878 & 0.261 & 0.447 &\\
SMC$\_$SC6  & 105318 & 0\uph52\upm30\zdot\ups84 & $-73\arcd13\arcm38\zdot\arcs4$ & 1.34489 & 0.061 & 5.893 & 14.598 & 0.077 & 0.337 &uncertain\\
SMC$\_$SC7  &  47305 & 0\uph54\upm55\zdot\ups60 & $-72\arcd44\arcm04\zdot\arcs1$ & 0.65083 & 0.000 &  --   & 16.939 & 0.300 & 0.469 &\\
SMC$\_$SC7  & 120135 & 0\uph55\upm14\zdot\ups35 & $-72\arcd45\arcm26\zdot\arcs8$ & 0.51687 & 0.000 &  --   & 17.267 & 0.267 & 0.441 &\\
SMC$\_$SC10 &  14774 & 1\uph03\upm37\zdot\ups13 & $-72\arcd32\arcm22\zdot\arcs7$ & 0.73343 & 0.000 &  --   & 16.335 & 0.282 & 0.452 &\\
SMC$\_$SC11 &  46599 & 1\uph07\upm30\zdot\ups54 & $-72\arcd38\arcm41\zdot\arcs2$ & 0.70284 & 0.000 &  --   & 16.772 & 0.291 & 0.469 &\\
\noalign{\vskip2pt}
\hline}

\setcounter{figure}{1}
%\begin{figure}[p]
%\vspace*{-3mm}
%\psfig{figure=fig1.ps,bbllx=100pt,bblly=150pt,bburx=460pt,bbury=730pt,clip=}
%\FigCap{Finding charts for single-mode second overtone Cepheid candidates.}
%\end{figure}

We were left with a list of 15 candidates for single-mode second
overtone   Cepheids which passed all four constraints: $R_{21}$ and
$\phi_{21}$ values,   CMD location and PL position. Table~1 contains
basic parameters of these   stars: ID, equatorial coordinates (J2000),
period in days, $R_{21}$ and   $\phi_{21}$ Fourier parameters and
intensity-mean {\it BVI} photometry.  Finding charts  for all candidates
are presented in Fig.~1. The size of the  {\it I}-band subframes is
${60\times60}$ arcsec; North is up and East to the  left. Fig.~2  shows
the phased light curves of all stars. {\it BVI} photometry   of all
objects will be available from the OGLE Internet archive when the  
catalog of Cepheids in the SMC is released. 

\begin{figure}[htb]
\psfig{figure=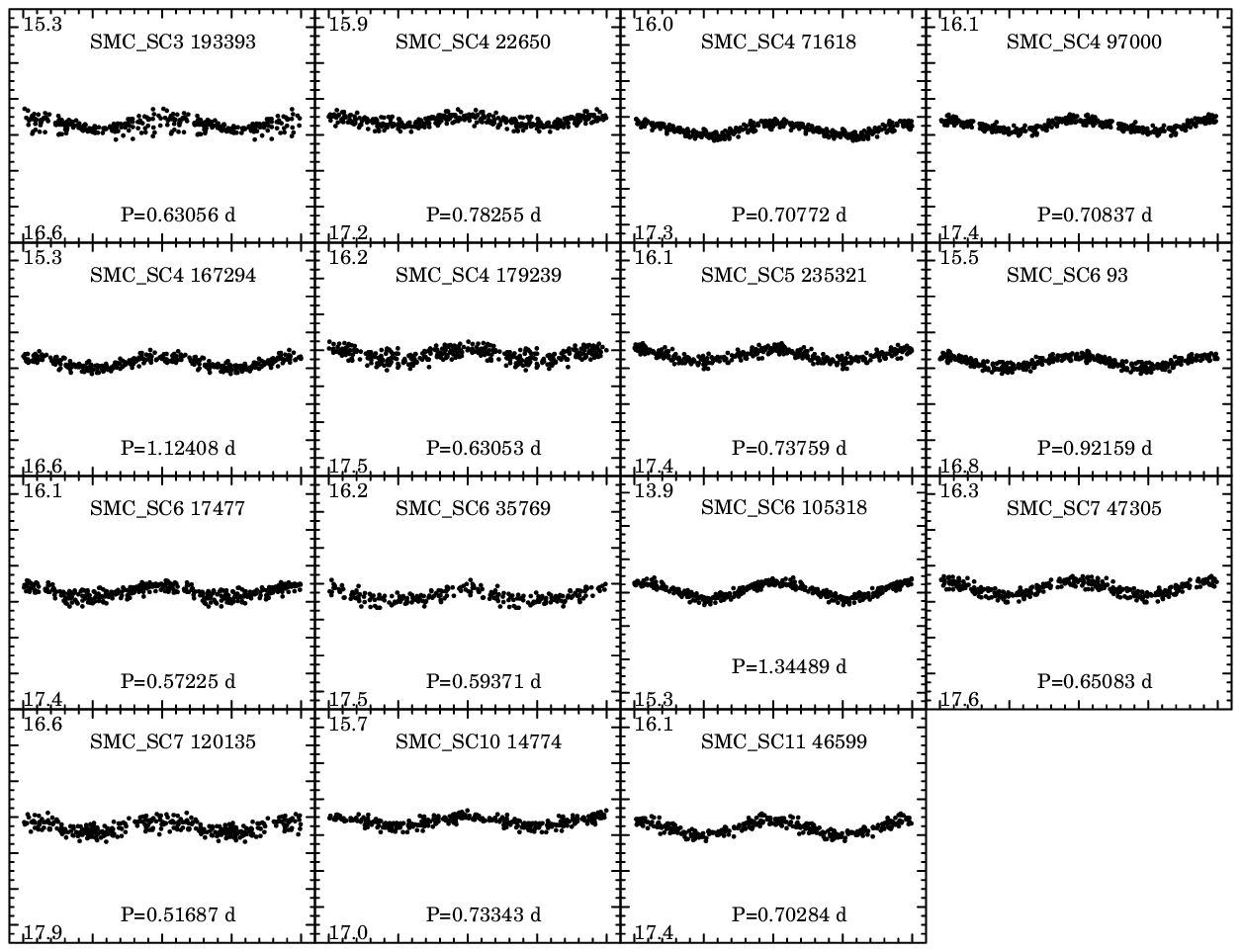,bbllx=120pt,bblly=310pt,bburx=490pt,bbury=595pt,width=12cm,clip=}
\vspace*{3pt}
\FigCap{Light curves of single-mode second overtone Cepheid candidates.}
\end{figure}

\begin{figure}[p]
\hglue-7mm{\psfig{figure=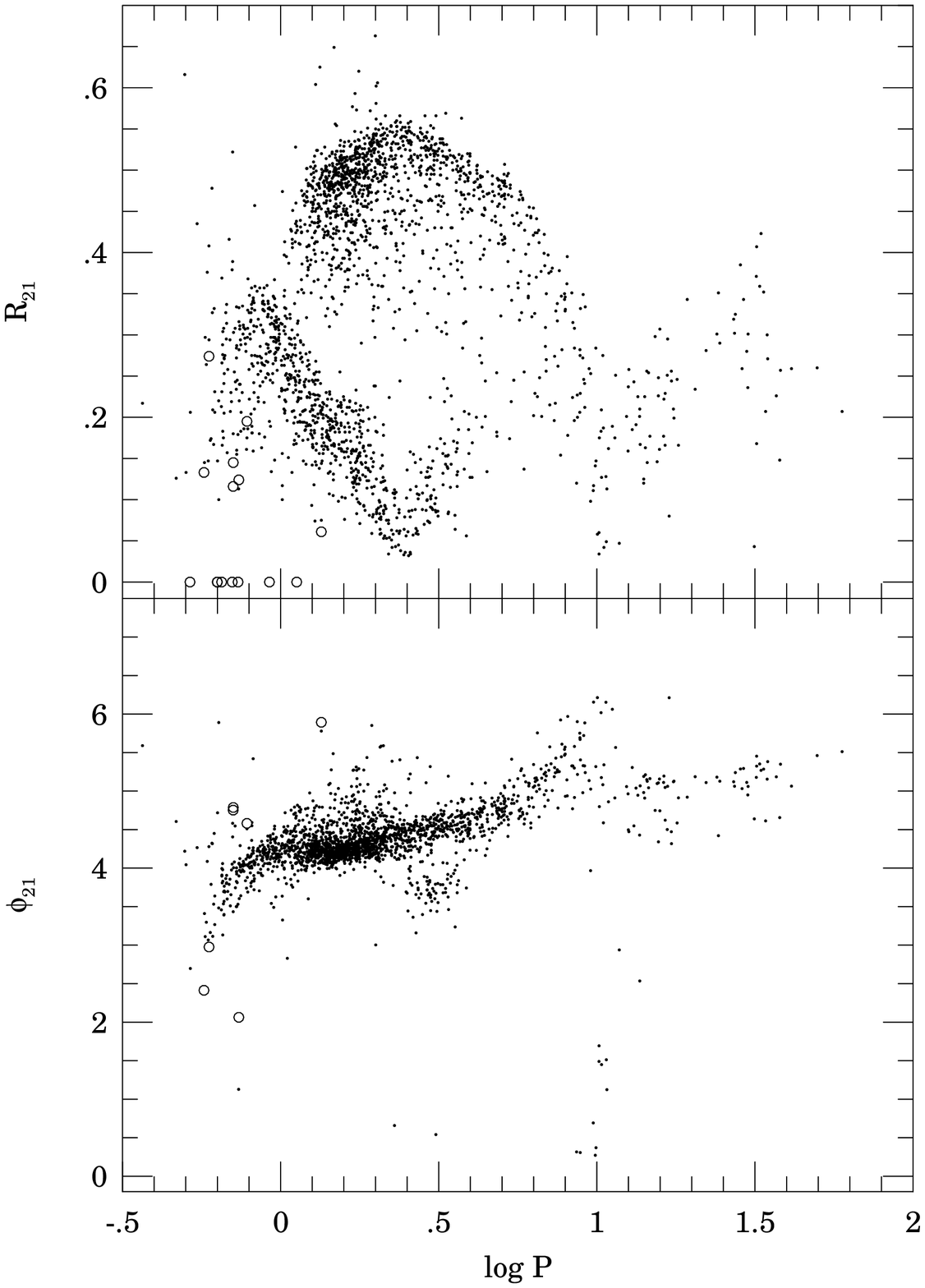,bbllx=25pt,bblly=40pt,bburx=505pt,bbury=705pt,width=13cm,clip=}}
\FigCap{${R_{21}}$ and ${\phi_{21}}$ \vs $\log P$ diagrams for single-mode
Cepheids from the SMC (small dots). Large open circles mark values of the 
second overtone Cepheid candidates.}
\end{figure} 
Fig.~3 shows $R_{21}$ \vs $\log P$ and $\phi_{21}$ \vs $\log P$ diagrams
with  positions of candidates marked by open circles. $R_{21}=0$ means 
non-significant value of this parameter, \ie sinusoidal light curve. In
this case  $\phi_{21}$ is not defined. 

\begin{figure}[p]
\hglue-7mm{\psfig{figure=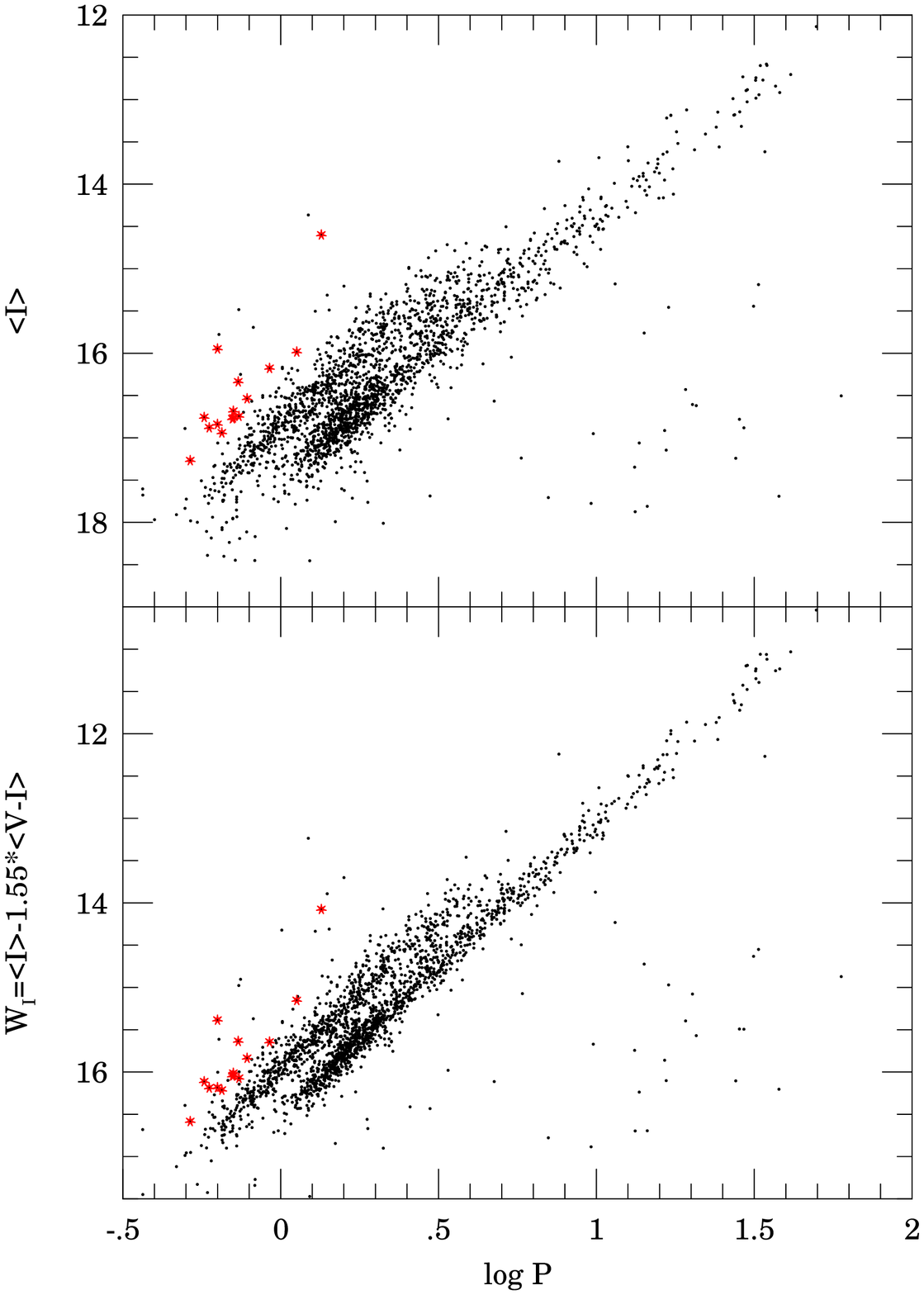,bbllx=25pt,bblly=40pt,bburx=505pt,bbury=710pt,width=13cm,clip=}}
\FigCap{Period-luminosity diagrams for Cepheids from the SMC. Star symbols 
mark positions of single-mode second overtone Cepheid candidates.}
\end{figure}
Fig.~4 presents PL diagrams with positions of candidates marked by star 
symbols. Abscissa in the upper panel of Fig.~4 is the intensity-mean 
{\it I}-band magnitude while in the lower panel the extinction
insensitive  magnitude $W_I=I-1.55\times(V-I)$. This  parameter, $W_I$,
removes effects  of interstellar extinction and partially color term of
the more general  period-luminosity-color relation for Cepheids. The
diagrams might still be  slightly contaminated by non-Cepheids as the
data come from the preliminary  version of the OGLE catalog of Cepheids
in the SMC. Nevertheless, the  diagrams, in particular that presented in
the lower panel of Fig.~4, are the  most tight and most complete PL
diagrams for classical Cepheids obtained so  far. 

It is difficult to assess completeness of the sample of second overtone 
candidates. The 16-17th mag objects with amplitude of 0.1~mag or below
are  close to the detection limit of the main OGLE variable star search.
The best  example are two candidates: SMC$\_$SC4 179239 and SMC$\_$SC6
35769 which are  located close to the edge of appropriate fields.
Therefore they should be also  detected in the overlapping fields but
they were not. In the case of  SMC$\_$SC4 179239 no periodicity was
detected in the counterpart photometry  due to somewhat larger errors of
measurements. In the second case -- the  standard deviation of
measurements of the counterpart of SMC$\_$SC6 35769 was  slightly below
the limit separating variable and non-variable stars and this  object
was not checked for variability. Nevertheless light curves of both 
counterpart stars  look almost identical like those of detected objects.
We may  conclude that more detailed reanalysis of the OGLE data of all
stars in the  Cepheid range of the CMD diagram, focused on low amplitude
variations can lead  to discovery of a few more candidates for
single-mode second overtone  Cepheids. 

\Section{Discussion}
15 candidates for single-mode second overtone Cepheids were found among
the  variable stars detected in the 2.4 square degree field in the
central part of  the SMC. This is the first so numerous sample of such
rare objects. Larger  population of the second overtone Cepheids was
predicted in lower metallicity  environment by theory and it seems that
this is indeed the case. Metallicity  of the SMC is smaller than that of
the LMC and Galaxy where only single  candidates were proposed. 

Are the selected objects indeed second overtone pulsators? The light
curves of  candidates presented in Fig.~2 show striking resemblance to
the light curves  of the second overtone pulsations in double-mode
Cepheids (\cf Appendix~C,  column~3 Udalski \etal 1999). In most cases
their light curves are almost  sinusoidal, the amplitude in the {\it
I}-band is about 0.1~mag. For many  candidates the $R_{21}$ parameter of
Fourier decomposition of the light  curve is statistically
non-significant indicating sinusoidal light curve  ($\phi_{21}$ is not
defined). For the remaining objects $R_{21}$ and  $\phi_{21}$ values
fall in similar regions where second overtone pulsation  values in
double-mode Cepheids can be found (Fig.~3 and Fig.~4 of Udalski  \etal
1999). 

Location of majority of objects in the PL diagrams (Fig.~4) is
consistent with  the expected location of the second overtone pulsators
induced from the second  to the first overtone period ratio in
double-mode Cepheids. Two objects,  however, namely SMC$\_$SC3 193393
and SMC$\_$SC6 105318, seem to be too bright  and therefore they are
uncertain candidates. The remaining candidates form a  sequence 
parallel to the sequences of the first overtone and fundamental mode 
Cepheids, but more quantitative approximation of the PL relation for
these objects  would be premature due to limited number of candidates. 

\begin{figure}[p]
\psfig{figure=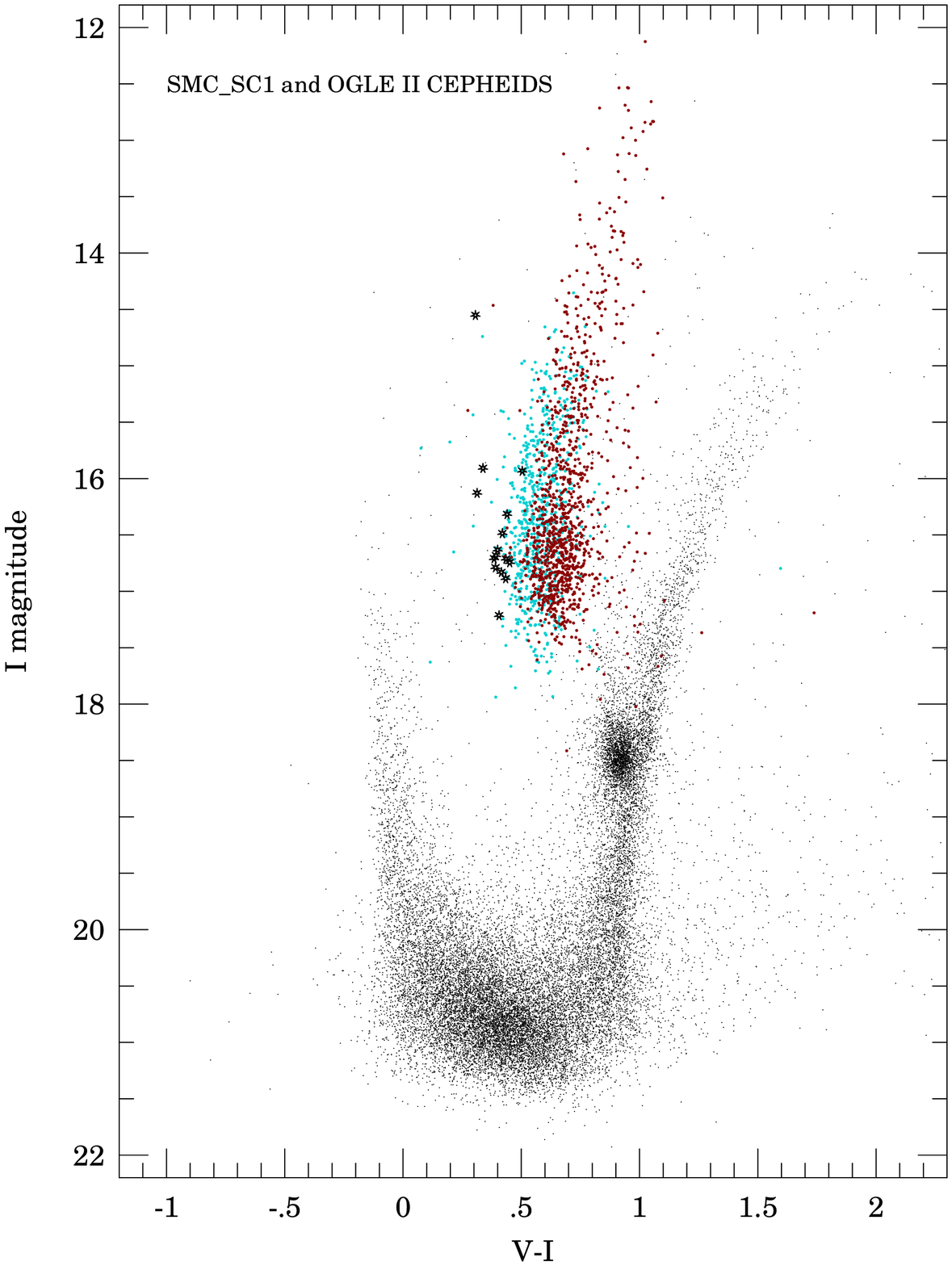,bbllx=35pt,bblly=50pt,bburx=550pt,bbury=730pt,width=13cm,clip=}
\FigCap{Color-magnitude diagram of the SMC$\_$SC1 field. Only about 20\% of field 
stars are plotted by tiny dots. Larger dots show positions of single-mode 
fundamental type Cepheids (darker dots) and first overtone stars (lighter 
dots). Star symbols mark positions of the single-mode second overtone Cepheid 
candidates.}
\end{figure}
Fig.~5 shows the color-magnitude diagram of the field SMC$\_$SC1. Only
about  20\% of stars from this field are plotted with tiny dots for
clarity. Larger  dots indicate positions of the first overtone (lighter
points) and fundamental  mode (darker points) Cepheids detected during
the OGLE search in 11 fields in  the SMC bar. The sample is corrected
for difference of the mean reddening   between SMC$\_$SC1 and other
fields as described in Udalski \etal (1999). The  reddening to the
SMC$\_$SC1 field is estimated to be ${E(V-I)=0.08}$ (Udalski  \etal
1999). 

Star symbols show positions of the single-mode second overtone Cepheid 
candidates in the CMD diagram. Magnitudes and colors of all these stars
were  also corrected for the mean difference of reddening between fields
and the  reference SMC$\_$SC1 field in the same way as for the remaining
Cepheids. 

One can notice that the second overtone Cepheid candidates  form a very 
homogeneous group as far as not only morphology of the light curves but
also  photometric properties are concerned. Except for three objects,
all the  remaining candidates have very similar colors and they
determine the  observational blue edge of the instability strip of the
first overtone  Cepheids. This is exactly the location where one would
expect second overtone  objects. The mean ${V-I}$ color of this group is
0.414, thus the stars would  be located at the end of the blue wing of
the distribution of $V-I$ color  indices of the first overtone Cepheids
in the SMC (\cf Fig.~6 Udalski \etal  1999). 

Two of three deviating objects in Fig.~5, namely SMC$\_$SC3 193393 and 
SMC$\_$SC6 105318, are the same stars which also deviate in the PL
diagram.  Their colors, in particular $B-V$, are too blue. Moreover, the
mean level of  brightness of both stars is changing by a few hundredths
of magnitude in the  time scale of months. Therefore variability of
these objects is very likely of  other origin, not related to
pulsations. The third deviating object SMC$\_$SC6  93 passes well the
remaining tests. Also its $B-V$ color is close  to the  colors of other
stars. Therefore we keep it in the list of single-mode  second  overtone
Cepheids candidates. 

Summarizing, the search for single-mode second overtone Cepheids in the
SMC  resulted in discovery of 13 firm candidates for the second overtone
pulsators.  They passed four photometric tests based on the Fourier
decomposition of the light  curve, location in the PL diagram and
location in the CMD diagram. Selected  stars form a very homogeneous
group with very similar light curves of almost  sinusoidal shape and
small amplitude. They have very similar color indices and  they
determine the observational blue edge of the instability strip of the 
first overtone Cepheids. All these facts strongly support the hypothesis
that  the stars are indeed the Cepheids pulsating solely in the second
overtone.  This is the first firm sample of such objects detected so
far. 

Spectroscopic observations could provide additional information
necessary to  prove that the objects from the presented sample are
indeed second overtone  pulsating Cepheids. All stars are relatively
bright and with the new large  telescopes in  southern hemisphere
starting operation in the coming years,  obtaining good resolution
spectroscopy of candidates should be feasible in the  near future.

\Acknow{We would like to thank Drs.\ W. Dziembowski and D.~Welch for
comments on the paper. The paper was partly supported by  the Polish KBN
grant 2P03D00814 to  A.\ Udalski.  Partial support for the OGLE  project
was provided with the NSF  grant AST-9530478 to B.~Paczy\'nski.}

\end{document}